\documentclass{article}
\usepackage{spconf,amsmath,graphicx}
\usepackage{amssymb}
\usepackage{multirow}

\usepackage{hyperref}
\hypersetup
{
	pdfauthor = UnnamedOrange,
	colorlinks = true,
	linkcolor = black,
	anchorcolor = black,
	citecolor = black,
	urlcolor = black
}

\title{FREETALKER: CONTROLLABLE SPEECH AND TEXT-DRIVEN GESTURE GENERATION BASED ON DIFFUSION MODELS FOR ENHANCED SPEAKER NATURALNESS}
\name{
 \normalsize
Sicheng Yang$ ^{1,*} $\thanks{$^{*}$Work performed during an internship at Tencent AI Lab.}, Zunnan Xu$ ^1 $, Haiwei Xue$ ^1$, Yongkang Cheng$ ^2 $, \textit{Shaoli Huang}$ ^3 $, \textit{Mingming Gong}$ ^{4,5} $, \textit{Zhiyong Wu}$ ^{1,\dagger} 
$\thanks{$ ^{\dagger} $Corresponding author.}
}

\address{
    $^1$ Tsinghua Shenzhen International Graduate School, Tsinghua University\ 
    $^2$ Northwest A\&F University \\
    $^3$ Tencent AI Lab\ 
    $^4$ University of Melbourne\ 
    $^5$ Mohamed bin Zayed University of Artificial Intelligence\\
    \small{yangsc21@mails.tsinghua.edu.cn, shaol.huang@gmail.com, mingming.gong@unimelb.edu.au, zywu@sz.tsinghua.edu.cn, }
}

\begin{document}

\maketitle

\begin{abstract}
% 目前的说话人大多是根据所说的音频和文本生成自发的演讲手势，没有进一步考虑说话人非说话的运动。
Current talking avatars mostly generate co-speech gestures based on audio and text of the utterance, without considering the non-speaking motion of the speaker. 
% Furthermore, previous works on co-speech gesture generation designed network structures on individual gesture dataset, resulting in limited data volume, generalizability, and limiting the speaker's movement.
Furthermore, previous works on co-speech gesture generation have designed network structures based on individual gesture datasets, which results in limited data volume, compromised generalizability, and restricted speaker movements.
To tackle these issues, we introduce FreeTalker, which, to the best of our knowledge, is the first framework for the generation of both spontaneous (e.g., co-speech gesture) and non-spontaneous (e.g., moving around the podium) speaker motions. % with enhanced controllability. 
% Specifically, we train a diffusion-based model for speaker motion generation, with unified representations of speech-driven gesture and text-driven motion utilizing heterogeneous data from various motion datasets. 
Specifically, we train a diffusion-based model for speaker motion generation that employs unified representations of both speech-driven gestures and text-driven motions, utilizing heterogeneous data sourced from various motion datasets.
%while extending the dataset. 
During inference, we utilize classifier-free guidance to highly control the style in the clips. Additionally, to create smooth transitions between clips, we utilize DoubleTake, a method that leverages a generative prior and ensures seamless motion blending. Extensive experiments show that our method generates natural and controllable speaker movements. 
% Our code, pre-trained model, and demo\footnote{Project url: https://youngseng.github.io/FreeTalker/} will be publicly available after the paper is accepted.
Our code, model, and demo are are available at \url{https://youngseng.github.io/FreeTalker/}.

\end{abstract}
\section{Introduction}
\label{sec:intro}

In various applications like virtual agents, animation, and human-computer interaction, the motions of a speaker are of paramount importance~\cite{nyatsanga2023comprehensive, kucherenko2023genea, zhuang2023gtn, xu2023chain}. These motions can be primarily divided into two segments: co-speech gestures that are inherently tied to the spoken content and non-spontaneous motions exhibited during talks~\cite{nyatsanga2023comprehensive, zhu2023human}.       % , such as interacting with the audience or navigating around the stage

% Speech-to-gesture is a task that generates human body gestures corresponding to speech input. The main challenges include generating high-quality and natural gestures, handling data scarcity, and effectively integrating multimodal input~\cite{nyatsanga2023comprehensive, kucherenko2023genea}. These difficulties make it a demanding problem in applications like virtual agents, animation, and human-computer interaction~\cite{nyatsanga2023comprehensive, zhuang2023gtn}.

In recent years, substantial focus has been dedicated to the generation of co-speech gestures. 
% Multimodal learning is widely adopted for co-speech gesture generation with style control. 
ZeroEGGS~\cite{ghorbani2023zeroeggs} emphasizes naturalness and zero-shot style control. 
\cite{alexanderson2023listen} adapts DiffWave for audio-driven motion synthesis, highlighting distinctive styles and control. DiffuseStyleGesture~\cite{yang2023diffusestylegesture} and GestureDiffuCLIP~\cite{ao2023gesturediffuclip} generate stylized gestures with exceptional human likeness and appropriateness.
% outperforming many previous approaches.
However, existing works primarily focus on global style control of co-speech gestures and do not facilitate free movement of the speaker, such as walking around the stage, pointing or looking in specific directions, or interacting with the audience. These aspects are crucial in presentations and speeches.
% Generating non-spontaneous movements of free speakers requires consistency with the input signal while conforming to human capabilities and physical laws~\cite{kong2023priority}. Recent works, such as 
In the domain of non-spontaneous motions~\cite{kong2023priority}, some works such as MDM~\cite{tevet2022human}, M2DM~\cite{kong2023priority}, and MotionDiffuse~\cite{zhang2022motiondiffuse}, have focused on text-controlled motion generation, achieving improvements in realism and controllability. PriorMDM~\cite{shafir2023human} introduces composition methods for denoising diffusion models. 

Despite these notable advancements, a significant gap remains.
To our knowledge, there hasn't been an effort that coherently integrates both of these motion categories. Challenges arise from varied motion representations, and multimodal learning intricacies.
MoFusion~\cite{ma2022pretrained} addresses dataset harmonization through pretraining for multi-task learning. Similarly, \cite{aberman2020skeleton} offers a framework for motion retargeting. UDE~\cite{zhou2023ude} introduces an engine for human motion sequences from diverse inputs.
UnifiedGesture~\cite{yang2023UnifiedGesture} employs further improvements in speech-driven gestures across multiple datasets. 
% To realize that it is challenging to utilize multiple datasets.
It's important to recognize the inherent challenges in utilizing multiple datasets.

In this paper, we propose a novel framework for generating both spontaneous and non-spontaneous speaker motions. Specifically, we first develop a diffusion-based model~\cite{ho2020denoising} for speaker motion generation, utilizing heterogeneous data from various motion datasets. Then, we employ classifier-free guidance \cite{ho2022classifier} during inference for highly controllable style in the generated clips. Additionally, we adopt DoubleTake \cite{shafir2023human} to create smooth transitions between clips and ensure seamless motion blending.
The main contributions of our work are:
(1) Proposing FreeTalker, the first framework to the best of our knowledge for generating both spontaneous and non-spontaneous speaker motions trained on multiple datasets.
(2) Incorporating classifier-free guidance and DoubleTake in our diffusion-based model for enhanced flexibility and control in gesture generation.
(3) Demonstrating improved naturalness in generated speaker motions through extensive experiments, surpassing existing approaches in terms of motion quality.

\section{Proposed Approach}
\label{sec:Method}

We aim to generate free-motion speakers using heterogeneous data from diverse motion datasets. 
In this section, we first describe the preprocessing steps required to integrate various motion data. 
Building on this, we introduce the diffusion model for motion generation. 
We then illustrate the controlled text-guided gesture generation method and explore long motion generation. Together, these components form a comprehensive system for effective generation of natural gestures.

\subsection{Motion Processing}
\label{sec:Motion Processing}

We expect that the features of the different motion datasets are correctly preserved.
In contrast to \cite{zhou2023ude} and \cite{yang2023UnifiedGesture}, where \cite{zhou2023ude} represents human motions with discrete codes and \cite{yang2023UnifiedGesture} retargets human motion to a homograph consisting of five terminal joints (head, hands, and feet), potentially losing important detailed information such as shoulders and fingers, our approach addresses this issue and preserves motion details.
We first convert the rotation matrix of the motion capture (BVH format) data to an axis angle representation of SMPL-X~\cite{pavlakos2019expressive}.
For the 3D position dataset, we fit it to the SMPL-X representation using VPoser~\cite{pavlakos2019expressive}.
We then scale the 3D translations of the root joint appropriately and adjust the initial orientation to be uniform across the different datasets as \cite{yang2023UnifiedGesture}.
With the SMPL-X model forward computation, we can obtain the 3D position of the SMPL-X representation. 
Then as in \cite{guo2022generating}, we use root height, root linear and rotational velocity, joint rotation, joint position, joint velocity, and foot contact as kinematic feature representations.
Each frame of the processed motion sequence has 659 dimensional features, which we denote as $\hat{x}_0 \in \mathbb{R}^{T_M \times 659}$, where $T_M$ denotes the number of motion sequence frames.

% \subsection{Architecture of the Proposed System}
\subsection{Diffusion Model for Motion Generation}

\begin{figure}
    \centering
    \includegraphics[width=0.99\linewidth]{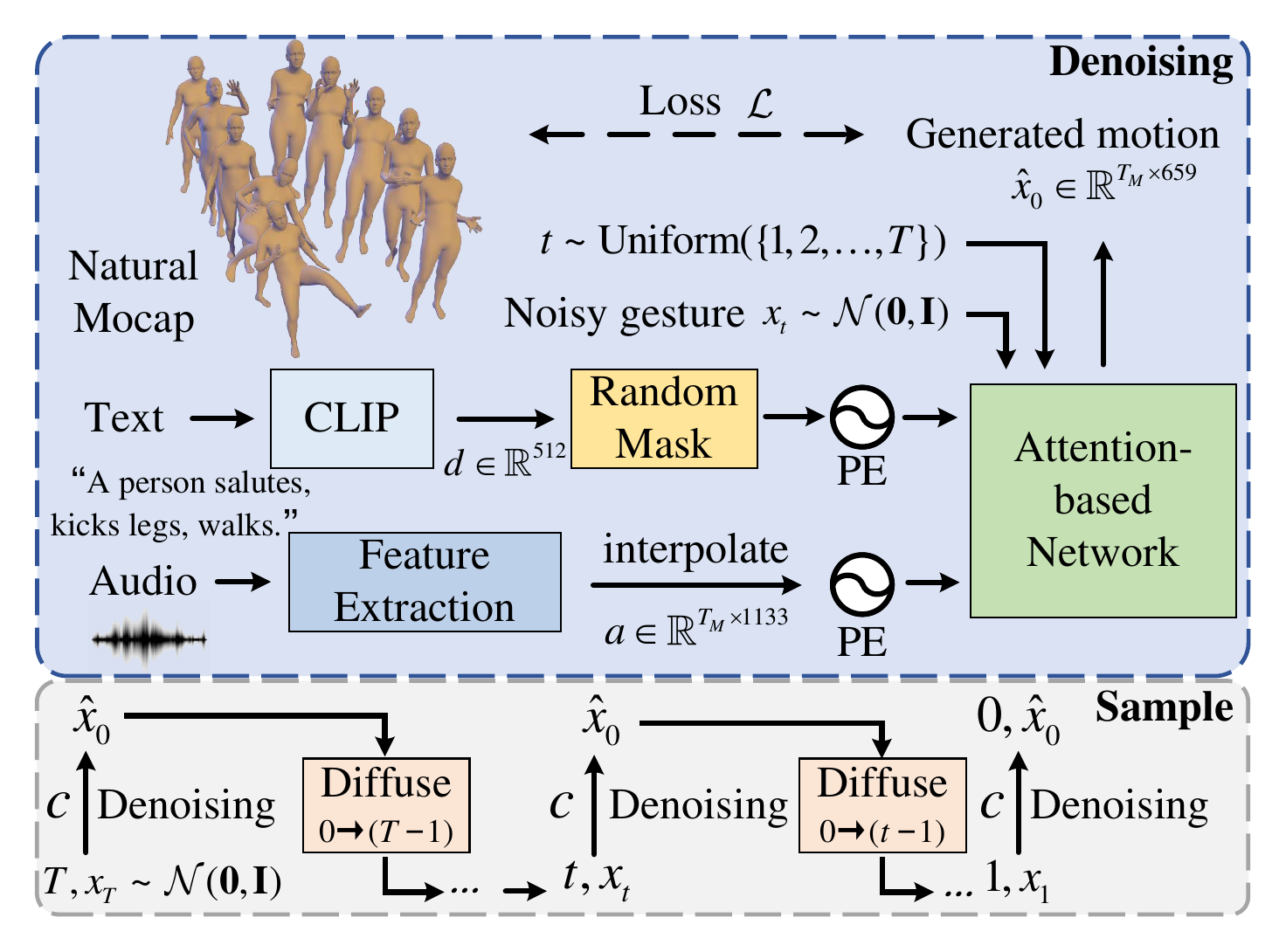}
    \caption{(Top) Denoising module.
   A noising step $t$ and a noisy motion sequence $x_t$ at this noising step conditioning on $c$ (including text description and audio) are fed into the model.
   PE indicates the addition of a positional encoding.
   (Bottom) Sample module.
   % At each noising step $t$, 
   We predict the $\hat{x}_0$ with the denoising process, then add the noise to the noising step $x_{t-1}$ with the diffuse process.
   This process is repeated from $t$ = $T$ until $t=0$.}
    \label{fig:diffusion}
\end{figure}

As illustrated in Figure~\ref{fig:diffusion}, we develop a diffusion model~\cite{ho2020denoising} inspired by \cite{tevet2022human} and \cite{yang2023diffusestylegesture}. For a noising step $t \in T$, we assume that $x_T \sim \mathcal{N}(0, I)$. The model assumes a stochastic process with $T$ noising steps:
$
q\left(x_t \mid x_{t-1}\right)=\mathcal{N}(\sqrt{\alpha_t} x_{t-1},(1-\alpha_t) I)
$.
The denoising process aims to predict the clean motion $\hat{x}_0$ from a noised motion $x_t$, a noise step $t$, a text condition encoded to CLIP~\cite{radford2021learning} space (represented by $d$, $d \in \mathbb{R}^{512}$), and an audio condition. The audio representation, consistent with \cite{yang2023diffusestylegesture+}, includes MFCC, Mel Spectrum, Pitch, Energy, WavLM \cite{chen2022wavlm}, and Onsets.
We perform linear interpolation of audio features in the time dimension to match the number of gesture frames, denoted as $a$, where $a \in \mathbb{R}^{T_M \times 1133}$. 
Subsequently, the textual description is spliced together as the first frame along with the speech embedding, the noise step, and the noisy action, feeding it into the self-attention~\cite{vaswani2017attention} layer, to yield the generated motion sequence.
The denoising process is expressed as $\hat{x}_0=Denoising\left(x_t, t, c\right)$, where $c = [d, a]$.
In practice, due to the lack of datasets with both non-spontaneous speaker motion and co-speech gestures, we blend datasets with speech-driven gestures and text-driven motions, and the missing modalities are set to zero during training.
The model is trained using Huber loss \cite{huber1992robust} function:
% combined with geometric losses that regulate joint position, rotation, velocity, and foot contact:
\begin{equation}
\mathcal{L}=E_{x_0 \sim q\left(x_0 \mid c\right), t \sim[1, T]}\left[\| x_0-\hat{x}_0 \|_2^2\right]
\end{equation}

During inference, at each noising step $t$, the original sample $\hat{x}_0$ is predicted and noised back to $x_{T-1}$. This process is iteratively repeated, starting from $t=T$ and continuing until $t=0$ is reached, resulting in more natural motion generation.

\subsection{Controllable text-guided gesture generation}

Generating gestures that are both expressive and consistent with textual descriptions is a challenge. Our diffusion model addresses this problem by extending the core idea of the classifier-free approach~\cite{ho2022classifier, alexanderson2023listen, yang2023diffusestylegesture} to adjust the strength of the non-spontaneous motion. As illustrated in Figure~\ref{fig:diffusion}, a random mask is added to the textual embedding for classifier-free learning. The classifier-free guidance of gesture generation is achieved by combining the predictions of the text-conditioned model $\operatorname{Denoise}\left(x_t, t, c_1\right)$, where $c_1 = [d, a]$, and the audio-conditioned model $\operatorname{Denoise}\left(x_t, t, c_2\right)$, where $c_2 = [ \varnothing, a]$, as follows:
\begin{align}
\label{8}
\hat{x}_{0, \gamma, c_1, c_2} &= \gamma \operatorname{Denoise}\left(x_t, t, c_1\right) + (1-\gamma) \operatorname{Denoise}\left(x_t, t, c_2\right)
\end{align}
where $\hat{x}_{0, \gamma, c_1, c_2}$ represents the combined output, and $\gamma$ is a parameter controlling the balance between the text-conditioned and audio-conditioned models.
In this work, the Denoising module learns both text-conditioned and audio-conditioned distributions by randomly masking 10\% of the samples using Bernoulli masks. 
% To control the style of the generated gestures, we leverage text descriptions $c$ in the condition. 
% When sampling, we can generate text-controlled gestures by interpolating or even extrapolating the two variants using $\gamma$. 
% We achieve this by setting $c_1 = [d_1, a]$ and $c_2 = [d_2, a]$ in Equation (\ref{8}). 
% This approach allows for greater flexibility and control the style of the generated gestures based on the given text descriptions.

\subsection{Long Motion Generation}

In tasks involving time series, a major challenge is generating long and coherent motion sequences.
Traditional approaches leveraging seed poses~\cite{yang2023diffusestylegesture, tseng2023edge} in generative tasks with non-time-aware sequences (e.g., text-to-motion) do not work well, so we use DoubleTake~\cite{shafir2023human} to generate long-distance motion.
Specifically, we first generate samples conditioned on its own text, audio and a handshake $\tau$ with its neighboring intervals through the denoising process, formulated as 
$
\tau_i=(1-\vec{\alpha}) \odot M_{i-1}[-h:]+\vec{\alpha} \odot M_i[: h]
$, 
where $h$ is the length of $\tau$, $M_i$ indicates the $i^{th}$ sequence $\alpha_j=j / h, \forall j: j \in[0: h)$ and $\odot$ indicates a element-wise multiplication.
% In the second stage of DoubleTake, 
Then we refine the transitions by reshaping the batch and focusing on the `transition sandwich' ($M_i$, $\tau_i$, $M_{i+1}$). We apply a soft-masking feature, using soft mask $M_{soft}$ and hard mask $M_{hard}$ for the sequence $M$ and handshake $\tau$. 
The masks ensure a gradual transition between the mask values, allowing $b$-frame-long linear masking between $M_{hard}$ and $M_{soft}$. This process refines the originally generated motion (suffix or prefix) to fit the transition during the second take at every denoising step. We partially noise and denoise the sandwich $T'$ noising steps:
$M'' = M' + M_{hard} \odot M_{soft} \odot (M'_{noisy} - M')$.
Here, $M''$ is the refined transition of the sequence, $M'$ is the original. 
Finally, we construct the long motion by unfolding the refined sequences and transitions, resulting in a smooth motion.

\section{Experiments}
\label{sec:Experiments}

\subsection{Experiment setup}

In our experiments, we use the text-driven (non-spontaneous) motion generation dataset HumanML3D~\cite{guo2022generating} and speech-driven (spontaneous) gesture generation dataset BEAT~\cite{liu2022beat}. 
% HumanML3D is a text-driven (non-spontaneous) motion generation dataset, while BEAT is a speech-driven (spontaneous) gesture generation dataset. 
All motion data are first resampled to 20 FPS.
For HumanML3D dataset, we only use data with motion frame counts between 40 and 180 frames, and the maximum text length for CLIP encoding is set to 20. During training, the motion sequence length is set to $T_M=180$ frames, with zero-padding for shorter sequences. 
And for BEAT, we use four English speaker gestures as described in \cite{liu2022beat} and randomly select a 180-frame segment of speech and corresponding gestures from a continuous gesture sequence.
To balance the number of motion data samples from both datasets, we employ weighted sampling to construct the dataloader.
% ensuring an equal distribution of data sources. 
All motion data are normalized by subtracting the mean and dividing by the standard deviation. 
The data is split into training, validation, and testing sets in an 8:1:1 ratio.
For the diffusion model, we use $T$ = 1000 noising steps and a cosine noise schedule.
The self-attention layer has a hidden space dimension of 256.
The batch size is 256, the learning rate is 2e-4, and the total number of training steps is set to 1 million. The model is trained on a V100 GPU for five days.
% For long-range motion generation, we use 
The DoubleTake method with a handshake size $h = 20$, a blend length $b=10$, a maximum $M_{hard}$ value of $85\%$, a minimum $M_{soft}$ value of $15\%$, and $T'=900$ denoising steps for $M'_{noisy}$.

\subsection{Experimental results and analysis}

\subsubsection{Visualization}

\begin{table*}[!htbp]
\centering
\caption{Quantitative results of comparison with the baseline models and ablation studies. `$\rightarrow$' denotes the closer to the real motion the better. `Naturalness' denotes the ``Ours vs. Compared model'' of the user study. `$^\star$' denotes ``Ours vs. Ground Truth'', implying a more rigorous evaluation, while entries without an asterisk are in reference to comparisons with other models. `$^*$' denotes the proposed model. `-' and `+' denote the removal and addition of the component, respectively.}
\label{tab:Ab}
\resizebox{\textwidth}{!}{%
\begin{tabular}{c|cccc|ccc|cc}
\hline
\multirow{2}{*}{Name} & \multicolumn{4}{c|}{Co-speech gesture generation} & \multicolumn{3}{c|}{Motion Generation} & \multicolumn{2}{c}{Free-motion} \\ \cline{2-10} 
                    & jerk $\rightarrow$             & acceleration $\rightarrow$  & FID $\downarrow$        & Naturalness $\uparrow$ & SSIM $\uparrow$  & FID $\downarrow$  & Naturalness $\uparrow$ & FID $\downarrow$   & Naturalness $\uparrow$ \\ \hline
Natural Mocap       & 135.36 $\pm$ 58.61  & 12.39 $\pm$ 11.79 & -          & -           & -     & -     & -           & -     & -           \\
DiffuseStyleGesture~\cite{yang2023diffusestylegesture} & 206.52 $\pm$ 83.65  & 5.68 $\pm$ 2.19   & 0.008      & 49\%        & -     & -     & -           & -     & -           \\
MDM~\cite{tevet2022human}                 & -       & -     & - & -           & 0.386 & 0.050 & 53\%        & -     & -           \\
Ours$^*$               & 245.78 $\pm$ 108.27 & 6.03 $\pm$ 2.55   & 0.139      & 40\%$^\star$       & 0.457 & 0.226 & 24\%$^\star$       & 0.139 & -           \\ \hline
\multicolumn{1}{l|}{\quad\quad\quad\quad- Huber loss$^*$}       & 226.30 $\pm$ 73.53  & 5.98 $\pm$ 2.33   & 0.027      & 52\%        & 0.389 & 0.041 & 53\%        & 0.029 & 54\%        \\ 
\multicolumn{1}{l|}{\quad\quad\quad\quad+ local attention$^*$}  & 203.77 $\pm$ 84.45  & 5.97 $\pm$ 2.51   & 0.005      & 49\%        & 0.431 & 0.051 & 54\%        & 0.032 & 52\%        \\ \hline
\end{tabular}%
}
\end{table*}

\begin{figure}[!tbp]
    \centering
    \includegraphics[width=0.99\linewidth]{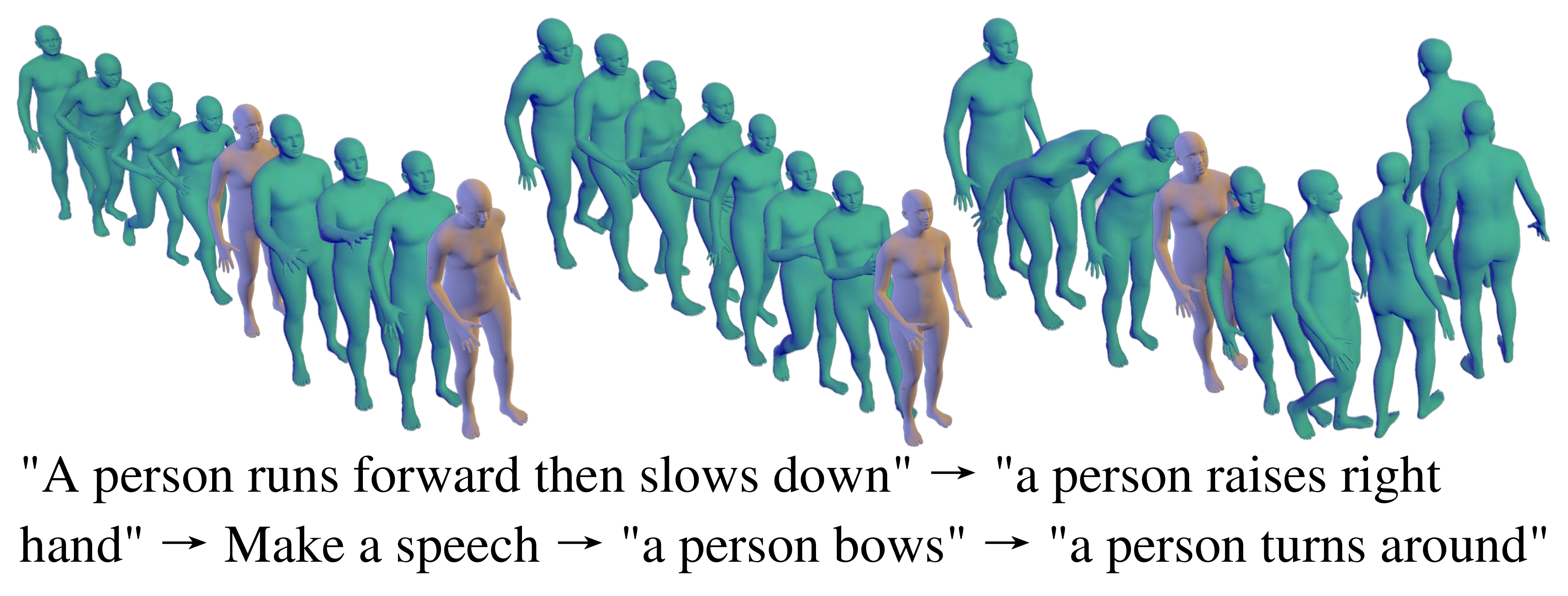}
    \caption{Visualization of FreeTalker generation. We can control the speaker's non-spontaneous motion through text, while the speaker generates spontaneous co-speech gestures from speech. The light yellow color indicates the model's ability to smoothly transition between motion segments.}
    \label{fig:vis-1}
\end{figure}

As illustrated in Figure~\ref{fig:vis-1}, FreeTalker generates a sequence of motions, including the speaker walking on stage, waving to the audience, delivering a speech, and finally bowing before leaving the stage. The generated motions exhibit smooth transitions between segments, allowing the speaker to move and speak in a natural manner.
\begin{figure}[!ht]
    \centering
    \includegraphics[width=0.825\linewidth]{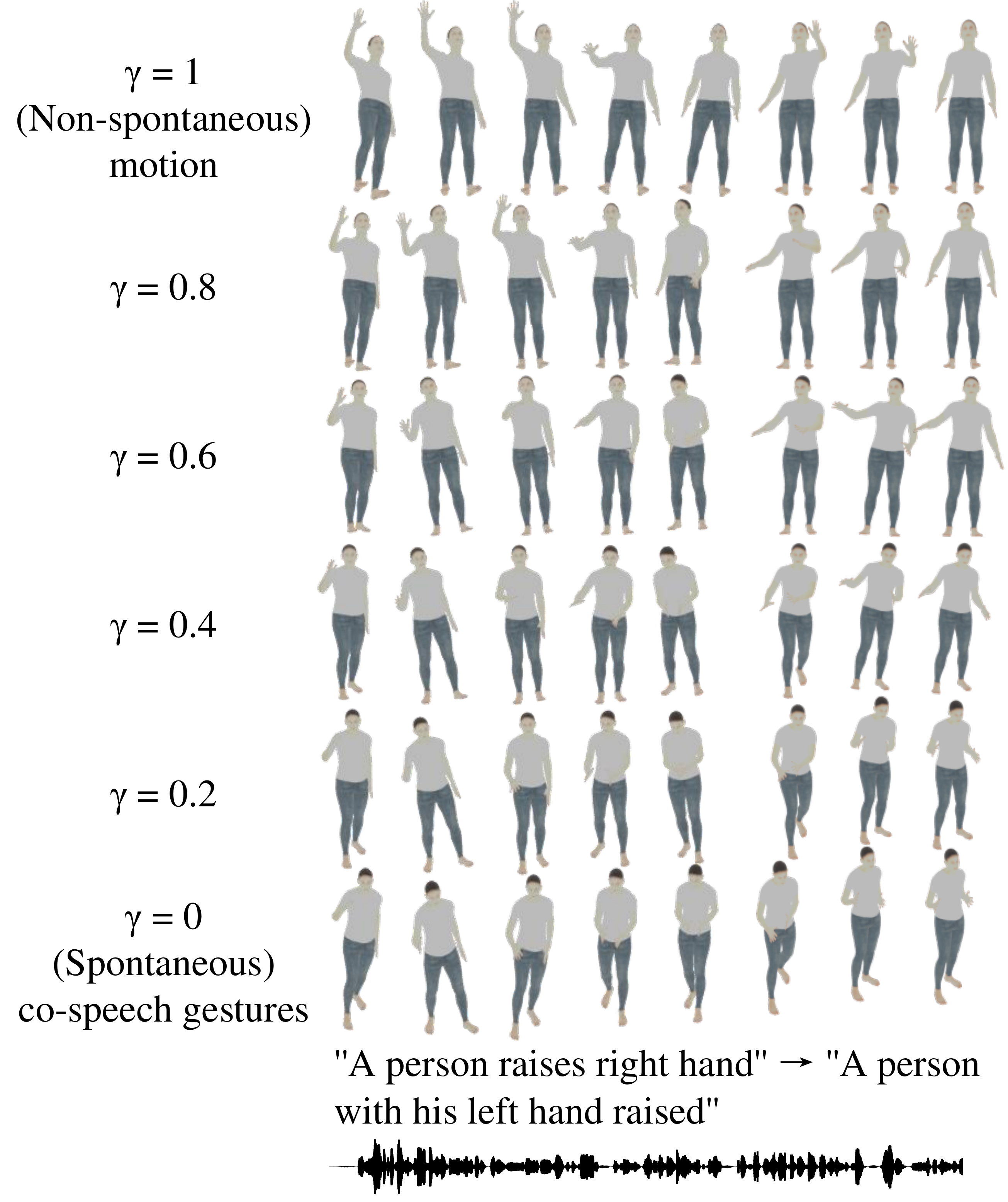}
    \caption{Visualization of style editing (non-spontaneous motion control) based on co-speech gestures. From top to bottom, generated motions gradually transition from text description-based control to spontaneous co-speech gestures based on speech, resulting in highly controllable gestures.}
    \label{fig:vis-2}
\end{figure}
As shown in Figure~\ref{fig:vis-2}, when $\gamma$ in Equation (\ref{8}) is set to 0, the gesture generation is conditioned only on speech input, enabling the model to produce co-speech gestures. As $\gamma$ gradually increases from 0 to 1, the model generates non-spontaneous gestures while maintaining alignment with speech. This allows us to freely edit the generated gestures and motions according to the text description.

\subsubsection{Objective Evaluation}

Due to the lack of methods capable of generating both spontaneous co-speech gestures and non-spontaneous motions, we evaluate each type of motion separately. 
% We select models~\cite{yang2023diffusestylegesture} and \cite{tevet2022human}, which have recently achieved excellent results, as our baseline models, respectively. 
We select \cite{yang2023diffusestylegesture} and \cite{tevet2022human} as our baseline models, as they have recently achieved excellent results.
% For co-speech gesture generation, we assess jerk, acceleration~\cite{kucherenko2019analyzing}, and FID~\cite{yoon2020speech}; while for textual description-driven motion generation, we evaluate SSIM \cite{hore2010image} and FID. 
For co-speech gesture generation, we assess jerk, acceleration~\cite{kucherenko2019analyzing}, and FID~\cite{yoon2020speech}; on the other hand, for textual description-driven motion generation, we evaluate SSIM \cite{hore2010image} and FID.
The results are shown in Table~\ref{tab:Ab}.
Our method attains competitive results with the baseline models for both generation tasks, demonstrating the effectiveness of our approach. Moreover, our method slightly outperforms the baselines in terms of jerk, acceleration and SSIM metrics.

\subsubsection{Subjective Evaluation}

To further evaluate the quality of the generated motions, we conducted a user study focusing on the naturalness (quality of the generated motions). 
% A total of 25 individuals participated in the scoring, evaluating our model against compared models. We presented the results of both models sequentially in the user interface and asked the users, "Which motion looks more human-like and more reasonable?" 
The study consisted of ten pairs of naturalness scoring, evaluating the naturalness of motions generated solely by co-speech gestures, solely by text-driven motions, and a combination of both.
During the evaluation, participants were presented with motion sequences generated by our model and the compared models. Following \cite{tevet2022human}, users were prompted with the question: "Which motion appears more human-like and reasonable?"
25 people participated in the study.
The results are shown in Table \ref{tab:Ab}. 
% It can be observed that our model demonstrates commendable performance, often rivaling the baseline models in terms of perceived naturalness, suggesting that expanding the motion database could further improve the performance.
A score closer to 100\% denotes higher naturalness.
It can be observed that our model demonstrates commendable performance, often rivaling the baseline models in terms of perceived naturalness. This suggests that expanding the motion database could further improve the performance.

Our method significantly enhances the Speech2Gesture and Text2Motion subtasks, as shown in Table \ref{tab:Ab}. It improves motion accuracy and naturalness, offering a diverse range of gestures, both spontaneous and non-spontaneous. This approach fills gaps in current methodologies and introduces a more adaptable motion generation framework.

\subsubsection{Ablation study}

To investigate the effectiveness of different components of our method, we designed the following ablation experiments. 
The results are detailed in the bottom two rows of Table \ref{tab:Ab}.
When the model is trained without Huber loss and instead uses MSE loss, the overall performance experiences a slight decline. Huber loss is more robust to outliers, generalizes better, and is better suited for smoothing the gradient to obtain a more coherent and natural sequence of actions. Furthermore, it converges to better results with fewer iterations. 
% When we feed $a$ into the local attention network~\cite{roy2021efficient} with relative position encoding~\cite{DBLP:conf/iclr/KitaevKL20} to extract the local information related to the gesture, placing it before the self-attention layer as suggested by \cite{yang2023diffusestylegesture}, the performance of co-speech gesture generation decreases slightly while the performance of motion generation improves, which illustrates the necessity of balancing different motion generation tasks to maintain optimal performance.
When we feed $a$ into the local attention network~\cite{roy2021efficient} with relative position encoding~\cite{DBLP:conf/iclr/KitaevKL20} to extract the local information related to the gesture before the self-attention layer as \cite{yang2023diffusestylegesture}, the performance of co-speech gesture generation decreases slightly. However, the performance of motion generation improves. This illustrates the necessity of balancing different motion generation tasks to maintain optimal performance.

\section{Conclusions}
\label{sec:Conclusions}

In this paper, we presented FreeTalker, a simple yet effective framework for generating both spontaneous and non-spontaneous speaker motions.  Leveraging a diffusion-based model, our method is trained on heterogeneous data sourced from various motion datasets.
The incorporation of classifier-free guidance and DoubleTake during inference stage ensures the natural, highly controllable and long-range motion generation.
Moreover, our approach lays the foundation for future work on large-scale motion datasets and more sophisticated models, paving the way for further advancements in speaker motion generation and enhancing talking avatars' naturalness in various applications.

We intend to elaborate on extending our work to the generation of fully digital humans, encompassing motions, facial expressions, and lip movements. We also aim to explore a more unified approach to digital human generation.

\section*{Acknowledgments}

This work is supported by National Natural Science Foundation of China (62076144), Shenzhen Key Laboratory of next generation interactive media innovative technology (ZDSYS20210623092001004), Shenzhen Science and Technology Program (WDZC20220816140515001, JCYJ20220-\\818101014030) and Tencent AI Lab Rhino-Bird Focused Research Program (RBFR2023015).

% \vfill\pagebreak

\bibliographystyle{IEEEbib}
\ninept
\bibliography{ICASSP2024}

\begin{thebibliography}{10}

\bibitem{nyatsanga2023comprehensive}
Simbarashe Nyatsanga, Taras Kucherenko, Chaitanya Ahuja, et~al.,
\newblock ``A comprehensive review of data-driven co-speech gesture
  generation,''
\newblock in {\em Computer Graphics Forum}, 2023, vol.~42, pp. 569--596.

\bibitem{kucherenko2023genea}
Taras Kucherenko, Rajmund Nagy, Youngwoo Yoon, et~al.,
\newblock ``The genea challenge 2023: A large scale evaluation of gesture
  generation models in monadic and dyadic settings,''
\newblock {\em arXiv preprint arXiv:2308.12646}, 2023.

\bibitem{zhuang2023gtn}
Haolin Zhuang, Shun Lei, Long Xiao, et~al.,
\newblock ``Gtn-bailando: Genre consistent long-term 3d dance generation based
  on pre-trained genre token network,''
\newblock in {\em IEEE International Conference on Acoustics, Speech and Signal
  Processing (ICASSP)}, 2023, pp. 1--5.

\bibitem{xu2023chain}
Zunnan Xu, Yachao Zhang, Sicheng Yang, Ronghui Li, and Xiu Li,
\newblock ``Chain of generation: Multi-modal gesture synthesis via cascaded
  conditional control,''
\newblock {\em arXiv preprint arXiv:2312.15900}, 2023.

\bibitem{zhu2023human}
Wentao Zhu, Xiaoxuan Ma, Dongwoo Ro, et~al.,
\newblock ``Human motion generation: A survey,''
\newblock {\em arXiv preprint arXiv:2307.10894}, 2023.

\bibitem{ghorbani2023zeroeggs}
Saeed Ghorbani, Ylva Ferstl, Daniel Holden, et~al.,
\newblock ``Zeroeggs: Zero-shot example-based gesture generation from speech,''
\newblock in {\em Computer Graphics Forum}, 2023.

\bibitem{alexanderson2023listen}
Simon Alexanderson, Rajmund Nagy, Jonas Beskow, et~al.,
\newblock ``Listen, denoise, action! audio-driven motion synthesis with
  diffusion models,''
\newblock {\em ACM Transactions on Graphics (TOG)}, vol. 42, no. 4, pp. 1--20,
  2023.

\bibitem{yang2023diffusestylegesture}
Sicheng Yang, Zhiyong Wu, Minglei Li, et~al.,
\newblock ``Diffusestylegesture: Stylized audio-driven co-speech gesture
  generation with diffusion models,''
\newblock {\em International Joint Conference on Artificial Intelligence},
  2023.

\bibitem{ao2023gesturediffuclip}
Tenglong Ao, Zeyi Zhang, and Libin Liu,
\newblock ``Gesturediffuclip: Gesture diffusion model with clip latents,''
\newblock {\em {ACM} Trans. Graph.}, 2023.

\bibitem{kong2023priority}
Hanyang Kong, Kehong Gong, Dongze Lian, et~al.,
\newblock ``Priority-centric human motion generation in discrete latent
  space,''
\newblock {\em arXiv preprint arXiv:2308.14480}, 2023.

\bibitem{tevet2022human}
Guy Tevet, Sigal Raab, Brian Gordon, et~al.,
\newblock ``Human motion diffusion model,''
\newblock in {\em The Eleventh International Conference on Learning
  Representations}, 2023.

\bibitem{zhang2022motiondiffuse}
Mingyuan Zhang, Zhongang Cai, et~al.,
\newblock ``Motiondiffuse: Text-driven human motion generation with diffusion
  model,''
\newblock {\em arXiv preprint arXiv:2208.15001}, 2022.

\bibitem{shafir2023human}
Yonatan Shafir, Guy Tevet, Roy Kapon, and Amit~H Bermano,
\newblock ``Human motion diffusion as a generative prior,''
\newblock {\em arXiv preprint arXiv:2303.01418}, 2023.

\bibitem{ma2022pretrained}
Jianxin Ma, Shuai Bai, and Chang Zhou,
\newblock ``Pretrained diffusion models for unified human motion synthesis,''
\newblock {\em arXiv preprint arXiv:2212.02837}, 2022.

\bibitem{aberman2020skeleton}
Kfir Aberman, Peizhuo Li, Dani Lischinski, et~al.,
\newblock ``Skeleton-aware networks for deep motion retargeting,''
\newblock {\em ACM Transactions on Graphics (TOG)}, vol. 39, no. 4, pp. 62--1,
  2020.

\bibitem{zhou2023ude}
Zixiang Zhou and Baoyuan Wang,
\newblock ``Ude: A unified driving engine for human motion generation,''
\newblock in {\em Proceedings of the IEEE/CVF Conference on Computer Vision and
  Pattern Recognition}, 2023, pp. 5632--5641.

\bibitem{yang2023UnifiedGesture}
Sicheng Yang, Zilin Wang, et~al.,
\newblock ``Unifiedgesture: A unified gesture synthesis model for multiple
  skeletons,''
\newblock {\em ACM International Conference on Multimedia}, 2023.

\bibitem{ho2020denoising}
Jonathan Ho, Ajay Jain, and Pieter Abbeel,
\newblock ``Denoising diffusion probabilistic models,''
\newblock {\em Advances in neural information processing systems}, pp.
  6840--6851, 2020.

\bibitem{ho2022classifier}
Jonathan Ho and Tim Salimans,
\newblock ``Classifier-free diffusion guidance,''
\newblock {\em arXiv preprint arXiv:2207.12598}, 2022.

\bibitem{pavlakos2019expressive}
Georgios Pavlakos, Vasileios Choutas, Nima Ghorbani, et~al.,
\newblock ``Expressive body capture: 3d hands, face, and body from a single
  image,''
\newblock in {\em Proceedings of the IEEE/CVF conference on computer vision and
  pattern recognition}, 2019, pp. 10975--10985.

\bibitem{guo2022generating}
Chuan Guo, Shihao Zou, Xinxin Zuo, et~al.,
\newblock ``Generating diverse and natural 3d human motions from text,''
\newblock in {\em Proceedings of the IEEE/CVF Conference on Computer Vision and
  Pattern Recognition}, 2022, pp. 5152--5161.

\bibitem{radford2021learning}
Alec Radford, Jong~Wook Kim, Chris Hallacy, et~al.,
\newblock ``Learning transferable visual models from natural language
  supervision,''
\newblock in {\em International conference on machine learning}. PMLR, 2021,
  pp. 8748--8763.

\bibitem{yang2023diffusestylegesture+}
Sicheng Yang, Haiwei Xue, Zhensong Zhang, et~al.,
\newblock ``The diffusestylegesture+ entry to the genea challenge 2023,''
\newblock {\em arXiv preprint arXiv:2308.13879}, 2023.

\bibitem{chen2022wavlm}
Sanyuan Chen, Chengyi Wang, Zhengyang Chen, et~al.,
\newblock ``Wavlm: Large-scale self-supervised pre-training for full stack
  speech processing,''
\newblock {\em IEEE Journal of Selected Topics in Signal Processing}, 2022.

\bibitem{vaswani2017attention}
Ashish Vaswani, Noam Shazeer, Niki Parmar, et~al.,
\newblock ``Attention is all you need,''
\newblock {\em Advances in neural information processing systems}, vol. 30,
  2017.

\bibitem{huber1992robust}
Peter~J Huber,
\newblock ``Robust estimation of a location parameter,''
\newblock in {\em Breakthroughs in statistics: Methodology and distribution},
  pp. 492--518. Springer, 1992.

\bibitem{tseng2023edge}
Jonathan Tseng, Rodrigo Castellon, and Karen Liu,
\newblock ``Edge: Editable dance generation from music,''
\newblock in {\em Proceedings of the IEEE/CVF Conference on Computer Vision and
  Pattern Recognition}, 2023, pp. 448--458.

\bibitem{liu2022beat}
Haiyang Liu, Zihao Zhu, Naoya Iwamoto, et~al.,
\newblock ``Beat: A large-scale semantic and emotional multi-modal dataset for
  conversational gestures synthesis,''
\newblock in {\em European Conference on Computer Vision}, 2022.

\bibitem{kucherenko2019analyzing}
Taras Kucherenko, Dai Hasegawa, Gustav~Eje Henter, Naoshi Kaneko, and Hedvig
  Kjellstr{\"o}m,
\newblock ``Analyzing input and output representations for speech-driven
  gesture generation,''
\newblock in {\em Proceedings of the 19th ACM International Conference on
  Intelligent Virtual Agents}, 2019, pp. 97--104.

\bibitem{yoon2020speech}
Youngwoo Yoon, Bok Cha, Joo-Haeng Lee, et~al.,
\newblock ``Speech gesture generation from the trimodal context of text, audio,
  and speaker identity,''
\newblock {\em ACM Transactions on Graphics (TOG)}, 2020.

\bibitem{hore2010image}
Alain Hore and Djemel Ziou,
\newblock ``Image quality metrics: Psnr vs. ssim,''
\newblock in {\em 2010 20th international conference on pattern recognition}.
  IEEE, 2010.

\bibitem{roy2021efficient}
Aurko Roy, Mohammad Saffar, Ashish Vaswani, et~al.,
\newblock ``Efficient content-based sparse attention with routing
  transformers,''
\newblock {\em Transactions of the Association for Computational Linguistics},
  vol. 9, pp. 53--68, 2021.

\bibitem{DBLP:conf/iclr/KitaevKL20}
Nikita Kitaev, Lukasz Kaiser, and Anselm Levskaya,
\newblock ``Reformer: The efficient transformer,''
\newblock in {\em International Conference on Learning Representations}, 2020.

\end{thebibliography}

\end{document}